\documentclass[aps,prd,reprint,twocolumn, groupedaddress,longbibliography,nofootinbib,10pt]{revtex4-1}
\usepackage{amsmath,amssymb}
\usepackage{graphicx}
\usepackage{color}
\usepackage{caption}
\usepackage{hyperref}
\usepackage{graphicx}
\usepackage[brazilian]{babel}
\usepackage[utf8]{inputenc}
\usepackage[T1]{fontenc}
\usepackage{perpage}
\MakePerPage{footnote}
\usepackage{hyperref}

\begin{document}
%opening
\title{Magnetic monopoles in Lorentz-violating electrodynamics\\[10pt]}

% \email[]{eslley@ift.unesp.br}
% \affiliation{SISSA, Via Bonomea 265, 34136 Trieste, Italy, \\
%      INFN sezione di Trieste, Via Valerio 2, 34127 Trieste, Italy}
% \author{Stefano Liberati}
% \email[]{stefano.liberati@sissa.it}
% \affiliation{SISSA, Via Bonomea 265, 34136 Trieste, Italy
%                                        and \\
%      INFN sezione di Trieste, Via Valerio 2, 34127 Trieste, Italy.}
\author{Rodrigo Turcati}\email{rodrigo.turcati@posgrad.ufsc.br}
\affiliation{Centro de Ciências Físicas e Matemáticas, Universidade Federal de Santa Catarina (UFSC), CEP 88040-900, Florianópolis, Santa Catarina, Brasil}
% \email[]{rturcati@sissa.it}
% \affiliation{UFSC, Universidade Federal de Santa Catarina.}
\author{Eslley Scatena}\email{e.scatena@ufsc.br}
\affiliation{Departamento de Ciências Exatas e Educação, Universidade Federal de Santa Catarina (UFSC), CEP 89065-300, Blumenau, Santa Catarina, Brasil}

\def\d{{\mathrm{d}}}
\newcommand{\scri}{\mathscr{I}}
\newcommand{\sun}{\ensuremath{\odot}}
\def\J{{\mathscr{J}}}
\def\L{{\mathscr{L}}}
\def\sech{{\mathrm{sech}}}
\def\T{{\mathcal{T}}}
\def\tr{{\mathrm{tr}}}
\def\diag{{\mathrm{diag}}}
\def\ln{{\mathrm{ln}}}
\def\Horava{Ho\v{r}ava}
\def\Aether{\AE{}ther}
\def\AEther{\AE{}ther}
\def\aether{\ae{}ther}
\def\UH{{\text{\sc uh}}} % small caps
\def\KH{{\text{\sc kh}}}

\begin{abstract}

% Magnetic monopoles may be introduced in a Lorentz
% In this paper we investigate

%Dirac-like magnetic monopoles are investigated in abelian Lorentz-violating electromagnetic models. 
%We investigate the possibility of accomodate Dirac-like monopoles in the framework of Lorentz-violating electrodynamics. To investigate this problem, we analyse three distinct models: Myers-Pospelov, Ellis et al. and  Gambini-Pullin. In the Myers-Pospelov case, which also violates the CPT symmetry, magnetic sources may be properly induced when it is accompanied by the appearance of an extra electric current. In addition, we also discuss the symmetry under duality transformations and the Dirac quantization condition in the aforementioned theories.

We investigate the possibility of coexistence between Dirac-like monopoles and Lorentz-violating electrodynamics. For this purpose, we study three distinct models: Myers-Pospelov, Ellis et al. and  Gambini-Pullin. In particular, we find that in the Myers-Pospelov electrodynamics, magnetic sources may be properly induced when it is accompanied by the appearance of an extra electric current. The symmetry under duality transformations and the Dirac quantization condition in the aforementioned theories is also discussed.
%In addition, we also discuss the symmetry under duality transformations and the Dirac quantization condition in the aforementioned theories.

%effective field theories... ultraviolet modifications of the dispersion relation...
%In this paper we present a discussion about the possibility of magnetic monopoles in higher-order Lorentz-violating electromagnetic theories. We consider three different Lorentz-violating models: the MP, GP and Ellis et al.

\end{abstract}

\maketitle

\section{Introduction}

Currently, there has been a growing interest in searching for small deviations of Lorentz invariance. Despite being one of the foundations of both general relativity and the standard model of particle physics, the recent availability of cosmological and astrophysical data, besides several terrestrial experiments, have increased the search for such symmetry violations (see e.g., \cite{Mattingly:2005re} for an extensive review). %Currently, the availability of cosmological and astrophysical data have been growing the in
%The recent availability of cosmological and astrophysical data, besides several earth-based experiments, have been 
%Currently, there has been a growing interest in searching Lorentz symmetry violations at high-energy regimes. One of the reasons for investigating deviations from the Lorentz invariance comes from the recent availability of cosmological and astrophysical data (see e.g.,\cite{Mattingly:2005re} for an extensive review). %Currently, the availability of cosmological and astrophysical data have been growing the interest in the searching of Lorentz symmetry violations at the high-energy regime (see e.g.,\cite{Mattingly:2005re} for an extensive review). 
From the theoretical standpoint, several preliminary computations in quantum gravity models suggest that Lorentz invariance might not be an exact symmetry of nature. Examples where such symmetry breaking may occur include string theory \cite{Kostelecky:1988zi}, loop quantum gravity \cite{Gambini:1998it}, non-commutativity of the spacetime \cite{Lukierski:1993wx,AmelinoCamelia:1999pm,Carroll:2001ws}, brane world scenarios \cite{Burgess:2002tb}, and so on \cite{AmelinoCamelia:1997gz,Barcelo:2005fc}. Besides that, ultraviolet divergences in local quantum field theory, a consequence of boost invariance of the field degrees of freedom, seems to indicate that Lorentz invariance should not be a fundamental symmetry at high energies \cite{Jacobson:2005bg}. 

%While the direct search for small deviations of the Standard Model of particle physics are expected to be supressed by the Planck scale, such tiny effects may, in principle, be magnified to measurable ones in the high energy regime \cite{Mattingly:2005re}. However, to describe this phenomenology, a complete theoretical formalism where Lorentz breakings take place is needed. In this vein, there has been a recent progress in theoretical physics by making use of effective field theories. For instance, Kostelecky and collaborators have developed a systematic framework that incorporate all possible Lorentz scalar operators at the lagrangian level that are responsible for the Lorentz and CPT symmetry violations \cite{Colladay:1998fq}. In addition, other prescriptions propose Lorentz breakings through the use of modified dispersion relations. The great advantage of making use of effective field theories is that one may acquire a quite good understanding of the physical process under consideration without knowing the underlying unified theory. 
 
In order to describe the phenomenology of these effects, a complete theoretical formalism where Lorentz breakings take place is needed. Among some proposals, %In this vein, 
there has been a recent progress in theoretical physics by making use of effective field theories. For instance, Kostelecky and collaborators have developed a systematic framework that incorporate all possible Lorentz scalar operators at the lagrangian level that are responsible for Lorentz symmetry violations \cite{Colladay:1998fq}. In addition, other prescriptions propose Lorentz breakings through the use of modified dispersion relations \cite{Jacobson:2002ye}. The great advantage of making use of effective field theories is the possibility of acquiring %that one may acquire 
a quite good understanding of the physical process under consideration without knowing the underlying unified theory. 

On the other hand, the theory of magnetic monopoles is of current interest in physics \cite{Vilenkin:2000jqa,Manton:2004tk,Balian:2005joa}. Despite never being observed in nature, several theoretical models propose their existence \cite{Goddard:1977da}. From the classical electromagnetism point of view, for instance, Maxwell equations become symmetric under duality transformation in the presence of electric and magnetic sources. In the domain of quantum mechanics, where electromagnetism is described in terms of potentials instead of fields, the situation is more intricate. To describe the electron motion under the influence of a magnetic field generated by a monopole alone, one needs to introduce a potential vector that is singular on a half-line, the so-called {\it Dirac string} \cite{Dirac:1931kp}. By imposing that such singularity does not be an observable, one derives the Dirac quantization condition, %Then, applying the standard canonical quantization will lead to the derivation of the Dirac condition, 
which states that electric charge discreteness follows from the existence of magnetic monopoles. Further developments in the field of monopoles have increased our understanding of several features in physics, ranging from a close connection between topology and physics \cite{Wu:1975es,Yang:1977bd} to grand unified theories \cite{tHooft:1974kcl,Polyakov:1974ek,Sorkin:1983ns,Gross:1983hb}. 

As is well known, the gauge invariance is crucial for the existence of magnetic poles \cite{Dirac:1931kp,Goddard:1977da}. With this aspect in view, the purpose of the present paper is to understand if Lorentz-violating electrodynamics that are gauge invariant and magnetic monopoles can coexist in the same scenario. %The theoretical comprehension of this feature may provide us new insights about the relevance of magnetic monopoles in theories that do not preserve Lorentz invariance. 
It is worth noting that previous works have investigated some of these issues in both abelian and non-abelian Lorentz-breaking theories \cite{Barraz:2007mi,BaetaScarpelli:2006mqz}. Here, we intend to extend the discussion to other Lorentz-violating electromagnetic models. 

The paper is organized as follows. In Sec. \ref{MPsection} we discuss if magnetic monopoles may be introduced in the Myers-Pospelov model. Sec. \ref{GPsection} and \ref{Ellissection} are devoted to the electrodynamics of Gambini-Pullin and Ellis et al., in this order. Features related to the duality symmetry and the Dirac quantization condition are investigated in Sec. \ref{dualitysection}. We present our conclusions in Sec. \ref{conclusion}.

In our conventions we use the Heaviside-Lorentz units with $\hbar=c=1$. The metric signature is $(+,-,-,-)$. 

% %***********************************************************************************************************

\section{Myers-Pospelov electrodynamics}\label{MPsection}

We start off our analysis by considering the photon sector of the Myers-Pospelov electrodynamics \cite{Myers:2003fd,Mariz:2011ed,Scatena:2014tha}, which is defined by the $U(1)$-gauge invariant lagrangian 
\begin{eqnarray}\label{MPlagrangian}
\mathcal{L}=-\frac{1}{4}F_{\mu\nu}F^{\mu\nu}+hn^{\mu}F_{\mu\nu}(n\cdot\partial)n_{\alpha}\tilde{F}^{\alpha\nu}, 
\end{eqnarray}
where $F_{\mu\nu}(=\partial_{\mu}A_{\nu}-\partial_{\nu}A_{\mu})$ is the field strength, $\tilde{F}^{\mu\nu}(\equiv\frac{1}{2}\epsilon^{\mu\nu\alpha\beta}F_{\alpha\beta})$ is the dual electromagnetic tensor, $A^{\mu}$ is the usual gauge field, and $n^{\mu}$ is a four-background vector that defines a preferred direction in spacetime. In addition, $h=\frac{\xi}{M_{P}}$ is the coupling constant, where $\xi$ is a dimensionless parameter supressed by the Planck mass $M_{P}$. To maintain the invariance under spacetime translations, $n^{\mu}$ is assumed to be constant. Together with the Lorentz invariance, CPT symmetry is also violated. The corresponding field equations, in the presence of an electric source $\mathcal{L}_{source}=-j_{\mu}A^{\mu}$, are
\begin{eqnarray}\label{MPequation}
&&\partial_{\mu}F^{\mu\nu}+2hn_{\alpha}\left(n\cdot\partial\right)^{2}\tilde{F}^{\nu\alpha}=j^{\nu}, 
%
%&&\label{Bianchiidentity}\partial_{\mu}\tilde{F}^{\mu\nu}=0.
\end{eqnarray}
whereupon $j^{\mu}=(\rho,\mathbf{j})$, where $\rho$ is the electric charge density and $\mathbf{j}$ is the electric current density. Besides the four-current $j^{\mu}$, let us suppose that there exist also a magnetic source $k^{\mu}=(\sigma,\mathbf{k})$, where $\sigma$ and $\mathbf{k}$ are the magnetic charge and the magnetic current, respectively. As a result, the Bianchi identity is not preserved, and takes the form
%Introducing now a magnetic source $k^{\mu}$, the Bianchi identity is not preserved, and one obtains
\begin{eqnarray}\label{MPmonopole}
\partial_{\mu}\tilde{F}^{\mu\nu}=k^{\nu}. 
\end{eqnarray}

An inspection of Eqs. (\ref{MPequation}) and (\ref{MPmonopole}) show us that Myers-Pospelov electrodynamics is not compatible with magnetic sources. %Now, it would be interesting to see if the above system is consistent. %To verify if the above system of equations is consistent, one may take the divergence of (\ref{MPequation}), which give us a nonconserved current 
%However, in the presence of magnetic sources, the MP field equations are not consistent. 
Thus, when one takes the divergence of Eq. (\ref{MPequation}), it follows that 
\begin{eqnarray}\label{nlconserved}
\partial_{\mu}j^{\mu}=2hn_{\mu}(n\cdot\partial)^{2}k^{\mu}, 
\end{eqnarray}
where the electric current $j^{\mu}$ is no longer conserved. 

At first sight, it seems that one cannot accomodate magnetic sources in the Myers-Pospelov model. One could be attempted to constraint the rhs of Eq. (\ref{nlconserved}) equals to zero. In such, since the four-background vector $n^{\mu}$ is constrained to be constant over all space, then, it would imply choose $n^{\mu}=0$. In this case, the five-dimensional operator would be trivial, and the Myers-Pospelov field equations would reduce to the standard Maxwell equations. Another possibility would be impose constraints on the magnetic source. In this context, purely spacelike $n^{\mu}=(0;\mathbf{n})$ and timelike  $n^{\mu}=(n^{0};\mathbf{0})$ configurations could, at least in principle, accomodate magnetic four-currents. However, these constraints seem to provide unphysical situations. For instance, let us consider a purely spacelike breaking in the radial direction, i.e., $n^{\mu}=(0,n^{r},0,0)$. As a result, Eq. (\ref{nlconserved}) reduces to
\begin{eqnarray}
\partial_{\mu}j^{\mu}=2hn_{r}^{3}\partial_{r}^{2}k^{r}. 
\end{eqnarray}

The above equation is satisfied for $k^{r}=r$, which seems to impose several restrictions in the way the magnetic currents could move around the space. Similar arguments could be used for Lorentz violations in other spatial directions. In the purely timelike case, in order to recover the electric charge conservation, the magnetic density charge $k^{0}$ could be, at most, linear in time. Although magnetic sources have never been observed in nature, such restrictions seem unlikely. Indeed, a full analysis of such scenario should be performed in order to disregard this possibility, which is out of the scope of this work. In the present paper, we will restrict our attention to general solutions for the magnetic source.

With this aspect in view, it was found in Ref. \cite{Barraz:2007mi}, by studing the CPT-odd Chern-Simons-type electrodynamics, that once the gauge symmetry is respected, magnetic sources may be properly introduced. Indeed, the presence of magnetic sources naturally induces the appearance of an extra electric current $j^{\mu}_{ind}$ in the Chern-Simons-type model. With this extra current, the total current becomes conserved, and one may introduce a magnetic four-current in the CPT-odd extension of Maxwell equations.

%On the other hand, in Ref. \cite{Barraz:2007mi}, by studing the CPT-odd Chern-Simons-type electrodynamics, it was shown that once the gauge symmetry is respected, magnetic sources may be properly introduced. Indeed, the presence of magnetic sources naturally induces the appearance of an extra electric current $j^{\mu}_{ind}$ in the Chern-Simons-type model. With this extra current, the total current becomes conserved, and one may introduce a magnetic four-current in the CPT-odd extension of Maxwell equations.

In what follows, we will assume that the same procedure may be applied in the context of the Myers-Pospelov electrodynamics. From this, the field equations (\ref{MPequation}) takes the form 
\begin{eqnarray}
\partial_{\mu}F^{\mu\nu}+2hn_{\alpha}\left(n\cdot\partial\right)^{2}\tilde{F}^{\nu\alpha}=j^{\nu}+j^{\nu}_{ind}, 
\end{eqnarray}
where the induced current satisfies the equation 
\begin{equation}\label{indcontinuityeq}
\partial_{\mu}j^{\mu}_{ind}=2hn_{\mu}(n\cdot\partial)^{2}k^{\mu}. 
\end{equation}

The above equation explicitly violates Lorentz symmetry under particle transformations, but preserve it under observer ones. Furthermore, the symmetry under discrete CPT transformations is violated. %In addition, $n^{\mu}$ violates discrete CPT symmetry. 

%, which can be written nonlocally as
% \begin{eqnarray}\label{nonlocallyinducedcharge}
% d_{t}q_{ind}+\oint_{S}\mathbf{j}_{ind}\cdot\mathbf{\hat{n}}da=h\int_{V}{d^{3}\mathbf{x}}n_{\mu}\left(n\cdot\partial\right)^{2}k^{\mu}.\nonumber\\
% \end{eqnarray}
Performing an integration over all space, Eq. (\ref{indcontinuityeq}) can be written nonlocally as %and assuming that $\mathbf{j}_{ind}$ goes to zero at spatial infinity, as usual, one gets
\begin{eqnarray}\label{indcurrent}
d_{t}q_{ind}=2h\int_{V}{d^{3}\mathbf{r}}n_{\mu}\left(n\cdot\partial\right)^{2}k^{\mu}.
\end{eqnarray}

On the other hand, the electric current induced by the appearance of the magnetic source turns out to be
%which show us that the induced charge is not globally conserved, as expected. Note that the induced four-current solution is 
\begin{equation}
j^{\mu}_{ind}=2hn_{\nu}(n\cdot\partial)^{2}\tilde{F}^{\mu\nu},
\end{equation}
or, equivalently, %in terms of its time and space components
\begin{equation}\label{inducedcurrentspaceandtime}
j^{\mu}_{ind}=2h(n\cdot\partial)^{2}\left(\mathbf{n}\cdot\mathbf{B};n^{0}\mathbf{B}-{\mathbf{n}\times\mathbf{E}}\right).
\end{equation}

%To check the consistency of the above outcomes, let us investigate the magnetic monopole configuration. Therefore

Now, let us analyse if our system is compatible with Dirac-like monopoles\footnote{The violation of Bianchi identity in Lorentz-violating electrodynamics leads, in general, to a class of solutions not restricted to magnetic ones. For this reason, we call such monopoles as Dirac-like objects.}. Adopting the static and pointlike magnetic pole, %Therefore, %Now, let us check the consistency of the above formalism. To do that, we will adopt the static and pointlike magnetic pole configuration 
$k^{\mu}=(g\delta^{3}(\mathbf{r}),\mathbf{0})$, where $g$ is the magnetic charge and gives $\mathbf{B}=(g/4\pi{r^{2}})\mathbf{\hat{r}}$, %and the corresponding Dirac delta is of the form $\delta^{3}\left(\mathbf{r}\right)=\delta\left(r\right)/4\pi{r^{2}}$, %Therefore, assuming spherical coordinates, 
it follows that Eq. (\ref{indcurrent}) assumes the form
\begin{eqnarray}\label{nonconservedtimelike}
d_{t}q_{ind}=2hn_{0}n_{r}^{2}\int_{V}{d^{3}\mathbf{r}}\frac{\partial^{2}}{\partial{r}^{2}}\left[g\delta^{3}(\mathbf{r})\right]. 
\end{eqnarray}

Note that the above relation depends uniquely on the timelike $n^{0}$ and the radial spacelike $n^{r}$ components. Therefore, we will neglect the angular dependence in the four-background vector $n^{\mu}$ in what follows. 

Using the corresponding Dirac delta in spherical coordinates, i.e.,  $\delta^{3}\left(\mathbf{r}\right)=\delta\left(r\right)/4\pi{r^{2}}$, then, Eq. (\ref{nonconservedtimelike}) is given by 
\begin{eqnarray}\label{indchargediverge}
d_{t}q_{ind}=24ghn_{0}n_{r}^{2}\int_{0}^{\infty}{d{r}}\frac{\delta(r)}{r^{2}}, 
\end{eqnarray}
which diverges at the point where the monopole is located. 

Let us also verify the consequences to the induced four-current. In the magnetic pole configuration, Eq. (\ref{inducedcurrentspaceandtime}) reads 
%Now, let us analyse if our system is compatible with Dirac-like monopoles. Adopting the static and pointlike magnetic pole, %Therefore, %Now, let us check the consistency of the above formalism. To do that, we will adopt the static and pointlike magnetic pole configuration 
%$k^{\mu}=(g\delta^{3}(\mathbf{r}),\mathbf{0})$, where $g$ is the magnetic charge and gives $\mathbf{B}=(g/4\pi{r^{2}})\mathbf{\hat{r}}$, %and the corresponding Dirac delta is of the form $\delta^{3}\left(\mathbf{r}\right)=\delta\left(r\right)/4\pi{r^{2}}$, %Therefore, assuming spherical coordinates, 
%it follows that Eq. (\ref{inducedcurrentspaceandtime}) assumes the form
\begin{eqnarray}\label{indtimelike}
%j^{\mu}_{ind}=n_{r}^{2}\nabla^{2}_{r}\left(\mathbf{n}\cdot\mathbf{B};n^{0}\mathbf{B}\right),\\
%
\rho_{ind}=\frac{3gh}{\pi}\frac{n_{r}^{3}}{r^{4}}, \quad\quad \mathbf{j}_{ind}=\frac{3gh}{\pi}n^{0}n_{r}^{2}\frac{\mathbf{\hat{r}}}{r^{4}},
\end{eqnarray}
where $\rho_{ind}$ and $\mathbf{j}_{ind}$ are the density charge and current, respectively.

Performing an integration over all space, the total induced charge blows up at the origin, say 
\begin{eqnarray}\label{indcharge}
&&q_{ind}=12ghn_{r}^{3}\int_{0}^{\infty}\frac{dr}{r^{2}}\rightarrow\infty. 
%
%&&I_{ind}=...
\end{eqnarray}

Furthermore, if a surface $S$ encloses the monopole charge $g$, the induced current $I_{ind}$ flowing through this surface is defined to be
\begin{eqnarray}\label{indsteady}
I_{ind}=\oint_{S}\mathbf{j}_{ind}\cdot\mathbf{\hat{n}}da=12ghn^{0}\frac{n_{r}^{2}}{r^{2}}, 
\end{eqnarray}
which again diverges at the monopole location.

Thus, for a generic external four-vector $n^{\mu}$, static and pointlike solutions leads to ill-defined quantities. On the other hand, considering the isotropic four-vector $n^{\mu}=(n^{0},\mathbf{0})$, the above system is trivially satisfied. i.e., 
%equations (\ref{indchargediverge}), (\ref{indcharge}) and (\ref{indsteady}) diverge. In this way, seems that spatially anisotropic , static and pointlike solutions leads to ill defined quantities. However, assuming that the four-background vector $v^{\mu}$ is purely timelike, then our system is trivially satisfied, i.e., 
\begin{eqnarray}
d_{t}q_{ind}=0, \quad \rho_{ind}=0 \quad {\it and} \quad \mathbf{j}_{ind}=\mathbf{0}.
\end{eqnarray}

The preceding analysis leads us to conclude that only in a preferred purely timelike direction one may properly introduce Dirac-like monopoles. % in the framework of the Myers-Pospelov electrodynamics. 

The isotropic model was the original proposal made by Myers and Pospelov \cite{Myers:2003fd} derived from the action
\begin{eqnarray}
S_{MP}=h\int{d^{4}x}\epsilon^{ijk}\dot{A}_{i}\partial_{j}\dot{A}_{k}. 
\end{eqnarray}

The above model modifies the Maxwell equations by introducing a higher-derivative term, which, in principle, could spoil the unitarity of the model. However, in Ref. \cite{Scatena:2014tha}, a fully analysis of the degrees of freedom in the electromagnetic sector of the Myers-Pospelov model have shown that the presence of ghosts in the purely timelike case can be avoided when one restricts the allowed values for the momenta. Indeed, since Myers-Pospelov electrodynamics is an effective field theory, one should expect that the energy range of the mentioned model is below the Planck scale. Such restriction is enough to ensure the unitarity for the purely timelike case. Therefore, we may conclude that magnetic charges can be safely accomodate in the isotropic configuration of the Myers-Pospelov electrodynamics.

On the other hand, recent astrophysical tests suggest that more general preferred background, such as space- and light-like cases, should be explored \cite{Reyes:2010pv}. However, in our prescription, static and magnetic pole solutions in the spatially anisotropic sector of the Myers-Pospelov electrodynamics introduce divergent quantities, which preclude the existence of magnetic charges in such configuration.

%the spatially anisotropic sector of the aforementioned model induce the appearance of divergences, which preclude the existence of magnetic charges in such background configuration.

In summary, magnetic sources may be accomodated in the Myers-Pospelov electrodynamics. On the other hand, only the purely timelike background configuration admits static and magnetic pole solutions. 
\section{Gambini-Pullin electrodynamics}\label{GPsection}

In the attempt to describe the light propagation in an emergent space-time arising from Loop Quantum Gravity perspective, it was found in Ref. \cite{Gambini:1998it} the appearance of a nonparity extension of the Maxwell electrodynamics. In this framework, the electric and magnetic fields, in the presence of external sources, are known to satisfy the equations 
\begin{eqnarray}\label{GPequations}
&\mathbf{\nabla}\cdot\mathbf{E}=\rho,& \\
&\mathbf{\nabla}\times\left(\mathbf{B}+2\chi\mathbf{\nabla}\times\mathbf{B}\right)-\partial_{t}\mathbf{E}=\mathbf{j},& \\
&\mathbf{\nabla}\cdot\mathbf{B}=\sigma,& \\
&\mathbf{\nabla}\times\left(\mathbf{E}+2\chi\mathbf{\nabla}\times\mathbf{E}\right)+\partial_{t}\mathbf{B}=-\mathbf{k},&
\end{eqnarray}
where the Lorentz violation is controlled by the parameter $\chi$. The above set of gauge invariant equations constitute the so-called Gambini-Pullin electrodynamics. It gives rise to a modified dispersion relation for the light propagation, which leads to Lorentz violations at high energies. %It is a gauge invariant theory. 
Furthermore, the continuity equation is conserved. Thus, contrary to the Myers-Pospelov electrodynamics, the above field equations do not require the introduction of an extra electric current.

Now, let us see if the above system is compatible with magnetic pole-like solutions. By taking the standard monopole configuration into account, we are left essentially with two equations for the magnetic field, say, %the field equations (\ref{GPequations}) reduce to 
\begin{eqnarray}
%&\mathbf{\nabla}\cdot\mathbf{E}=0,& \\\label{rotBGP}
&\label{divBGP}\mathbf{\nabla}\cdot\mathbf{B}=g\delta^{3}(\mathbf{r}),& \\
&\label{rotGP}\mathbf{\nabla}\times\left(\mathbf{B}+2\chi\mathbf{\nabla}\times\mathbf{B}\right)=\mathbf{0},& 
%&\mathbf{\nabla}\times\left(\mathbf{E}+2\xi\mathbf{\nabla}\times\mathbf{E}\right)=\mathbf{0}.&
\end{eqnarray}
which have the magnetic field solution $\mathbf{B}=(g/4\pi{r^{2}})\mathbf{\hat{r}}$. Indeed, the Dirac-like magnetic field satisfies both equations (\ref{divBGP}) and (\ref{rotGP}), and magnetic monopoles can be safely accomodate in the framework of the Gambini-Pullin electrodynamics.  

%Thus, we may conclude that magnetic monopoles can be safely accomodate in the framework of the Gambini-Pullin electrodynamics.  

%According to the Helmholtz theorem, a vector field that has both divergent and rotational null, and goes to zero at spatial infinity, is identically null. Therefore, the electric field $\mathbf{E}$ is zero. Furthermore, the radial magnetic field generated by the monopole is $\mathbf{B}=\frac{g}{4\pi}\frac{\mathbf{\hat{r}}}{r^{2}}$. Note that the magnetic field $\mathbf{B}$ solution satisfies both (\ref{rotBGP}) and (\ref{divBGP}), which show us that one may incorporate magnetic monopoles in the Gambini-Pullin theory. 

%*********************************************************************************
\section{Ellis et al. electrodynamics}\label{Ellissection}

Up to now, string theory is the only consistent formulation that incorporate gravity along with the other fundamental interactions. In Refs. \cite{AmelinoCamelia:1996pj,Ellis:1999uh,Ellis:1999jf,Ellis:1999sd,Ellis:1999sf,Ellis:2003ua,Ellis:2003sd}, by using the Liouville approach to noncritial string theory, it was investigated the effects of quantum gravitational vacuum fluctuations on the light propagation. Among some results, it was found that the photon propagation through the spacetime foam lead to a subluminal dispersion relation. Thus, in the Ellis et al. scenario, electric and magnetic fields in the presence of electromagnetic sources satisfy the equations %In terms of the field strengths $\mathbf{E}$ and $\mathbf{B}$, the Ellis et al. field equations read
\begin{eqnarray}
&\mathbf{\nabla}\cdot\mathbf{E}+\mathbf{u}\cdot\partial_{t}\mathbf{E}=\rho_{eff},& \\
&\mathbf{\nabla}\times\mathbf{B}-(1-u^{2})\partial_{t}\mathbf{E}+\mathbf{u}\times\partial_{t}\mathbf{B}+(\mathbf{u}\cdot\mathbf{\nabla})\mathbf{E}=\mathbf{j}_{eff},& \\
&\mathbf{\nabla}\cdot\mathbf{B}=\sigma,& \\
&\mathbf{\nabla}\times\mathbf{E}+\partial_{t}\mathbf{B}=-\mathbf{k},&
\end{eqnarray}
where in the momentum space $\mathbf{u}=f(w)\mathbf{k}$, $\rho_{eff}=\rho-\mathbf{u}\cdot\mathbf{j}$ and $\mathbf{j}_{eff}=\mathbf{j}+\mathbf{u}(\rho-\mathbf{u}\cdot\mathbf{j})$ are the effective electric density and currents, respectively. The vector $\mathbf{u}$ leads to a breakdown of the Lorentz invariance since it is related to the recoil velocity of the photon \cite{AmelinoCamelia:1996pj,Ellis:1999uh,Ellis:1999jf,Ellis:1999sd,Ellis:1999sf,Ellis:2003ua,Ellis:2003sd}. 

We may now investigate the possibility of magnetic pole solutions. Therefore, adopting the static and pointlike magnetic charge configuration, the resulting system of field equations assume the form
\begin{eqnarray}
&\mathbf{\nabla}\cdot\mathbf{E}=0,& \\
&\mathbf{\nabla}\times\mathbf{B}+(\mathbf{u}\cdot\mathbf{\nabla})\mathbf{E}=\mathbf{0},& \\
&\mathbf{\nabla}\cdot\mathbf{B}=g\delta^{3}(\mathbf{r}),& \\
&\mathbf{\nabla}\times\mathbf{E}=\mathbf{0}.&
\end{eqnarray}

Note that the divergent and rotational of the electric field are both zero, which, according to the Helmholtz theorem, give us $\mathbf{E}=\mathbf{0}$. Thus, the above configuration reduces to the Maxwell equations in the presence of magnetic charges in the absence of electric fields. 

As well as in the Gambini-Pullin model, the Ellis et al. electrodynamics provide a suitable scenario to incorporate magnetic monopoles. %, which is compatible with magnetic pole solutions. 

%Therefore, the Ellis et al. electrodynamics is compatible with magnetic charges.
%***************************************************************************
\section{Duality symmetry and the Dirac quantization condition}\label{dualitysection}

We shall now discuss the role of the duality symmetry and the Dirac quantization condition for the above Lorentz-violating electrodynamics. 

\subsection{Duality symmetry}

In the absence of sources, Maxwell equations are symmetric under the duality transformation 
\begin{eqnarray}\label{dualitysymmetry}
F^{\mu\nu}\rightarrow\tilde{F}^{\mu\nu}, \quad \tilde{F}^{\mu\nu}\rightarrow-{F^{\mu\nu}}.
\end{eqnarray}

However, the introduction of an electric four-current $j^{\mu}$ violates such symmetry. In order to recover the duality invariance of Maxwell theory, one needs to add a magnetic source $k^{\mu}$. Hence, both electric and magnetic four-currents satisfy the transformation
\begin{eqnarray}\label{dualitycurrent}
j^{\mu}\rightarrow{k}^{\mu}, \quad {k}^{\mu}\rightarrow-{j^{\mu}},
\end{eqnarray}
and the invariance under duality transformations is fully restored, which ensures that the standard electromagnetic results involving electric charges may be translated to magnetic ones.

With regard to the theories under investigation, only Gambini-Pullin preserve the duality invariance. %the duality symmetry in the presence of electric and magnetic sources. 
Thus, unlike Gambini-Pullin electrodynamics, the violation of Lorentz symmetry in the Myers-Pospelov and Ellis et al. models show us that there is an asymmetry between electric and magnetic fields even in the absence of electromagnetic sources. This is due to the fact that both Myers-Pospelov and Ellis et al. models have their dynamics modified by the presence of an extra field. On the other hand, in the Gambini-Pullin framework, the dynamics is changed due the appearance of higher-derivatives in both electric and magnetic fields parametrized by a parameter that is linked to the Lorentz violation. Indeed, the modification introduced in this framework maintain the symmetry under duality transformations. %of a new parameter responsible for the Lorentz violation. This parameter comes together with both electric and magnetic fields in a symmetric way, preserving the duality invariance of the model.

%In such theories, the field responsible for the Lorentz violation induce an asymmetry between electric and magnetic fields even in absence of sources the vacum

\subsection{Dirac quantization condition}

Another interesting aspect of magnetic monopoles is related to the Dirac quantization condition. As is well known, the existence of the Dirac monopole is linked to the gauge invariance of the corresponding theory. More precisely, the quantum mechanical description of the electron under the influence of a magnetic field generated by a monopole leads, by following the standard methods, to the condition
\begin{eqnarray}\label{diracondition}
qg=2\pi{n}, 
\end{eqnarray}
where $q$ is the electric charge and $n$ is an integer. This is the so-called Dirac quantization condition. It is important to note that condition (\ref{diracondition}) emerge by the fact that for any closed surface contaning a magnetic charge $g$, we have  
\begin{eqnarray}\label{magneticmonopoleequation}
\oint_{S}\mathbf{B}\cdot\mathbf{\hat{n}}da=g.
%\mathbf{\nabla}\cdot\mathbf{B}=0. 
\end{eqnarray}

According to Eq. (\ref{magneticmonopoleequation}), in order to write the magnetic field $\mathbf{B}$ in terms of the potential vector $\mathbf{A}$, %Eq. (\ref{magneticmonopoleequation}) tell us that in order to write the magnetic field in terms of the potential vector, 
one needs to introduce the notion of {\it Dirac string}. Indeed, when one imposes that such object is not an observable, then the Dirac quantization condition (\ref{diracondition}) is obtained (For further details, see Ref. \cite{Goddard:1977da}).

From the preceding considerations, we come to conclusion that both Gambini-Pullin and Ellis et al. electrodynamics preserve the Dirac condition. Thus, these models have the standard magnetic pole solution $\mathbf{B}=(g/4\pi{r^{2}})\mathbf{\hat{r}}$, which ensure the existence of the Dirac quantization condition. Magnetic monopoles in the Myers-Pospelov framework, in turn, can be properly introduced when the model has a timelike preferred direction. In such, the Dirac condition is acquired. %This is the only configuration that preserves the electric charge discreteness. 
On the other hand, as discussed in Sec. \ref{MPsection}, space- and light-like directions do not provide a consistent scenario to incorporate magnetic charges in the Myers-Pospelov electrodynamics.

%This is the background vector 

%Lorentz-violating models under investigation preserve the Dirac condition (\ref{diracondition}). This result can be viewed as a direct consequence of the fact that all models satisfy Eq. (\ref{magneticmonopoleequation}), except in the space- and light-like cases of the Myers-Pospelov model. Indeed, as discussed in Sec. \ref{MPsection}, one cannot even introduce magnetic monopole-like solutions in these background configurations.

\section{Final Remarks}\label{conclusion}

%total flux through a sphere surrounding the origin is ...

%abelian level of electrodynamics... total induced current... Indeed, the extra electric current induces the appearance of a new charge, which is not globally conserved.

%properties of the induced charge... let us pay attention to the electric current induced by the appearance of this monopole in the MP framework. %From eqs. it follows that

%violate Lorentz symmetry under particle transformation while preserve the observer ones.

In this paper we have analyzed the possibility of coexistence between magnetic monopoles and three known Lorentz-violating abelian electrodynamics, namely, Myers-Pospelov, Ellis et al. and Gambini-Pullin. In the Myers-Pospelov model we found that Dirac-like monopole may be introduced provided it is accompained by the addition of an extra electric current. This mechanism was possible since the gauge symmetry is preserved. %Myers-Pospelov electrodynamics is a gauge-invariant model. 
Furthermore, the existence of magnetic poles also depend on the choice of the four-vector $n^{\mu}$. For the purely timelike case, the induced current solutions are trivially satisfied and magnetic poles may be properly introduced. The purely space- and light-like breakings in the presence of magnetic charges, in turn, lead to inconsistencies. 

As an aside, we would like to remark that although Dirac-like monopoles cannot be accomodated in the space- and light-like preferred directions in the present prescription, maybe another method might be suitable to introduce such objects. A possible scenario could be a supersymmetric extension of the Myers-Pospelov model. In such, the emergence of new terms could eventually cancel out the divergences and lead to well-defined quantities. %On the other hand, the gauge invariance of the Myers-Pospelov model is preserved for arbitrary spacetime directions. %even in the space- and light-like configurations, the Myers-Pospelov model remains gauge invariant. 
%It seems to indicate us that maybe another mechanism should be possible in order to accomodate magnetic monopoles in the space- and light-like  backgrounds. One possibile scenario could be a supersymmetric extension of the Myers-Pospelov, which could bring new terms that could eventually cancel out the divergences. 
Another way to get rid of these infinities could be by performing a spatial projection of the Myers-Pospelov electrodynamics in a reduced dimensional model similar to what have been made in the Chern-Simons-type model \cite{Barraz:2007mi}.

With regard to the duality symmetry, we have found that only Gambini-Pullin electrodynamics preserve it. Actually, even in the absence of sources, Myers-Pospelov and Ellis et al. models are not invariant under these transformations. Indeed, the duality symmetry is a peculiarity of the electromagnetism in four dimensions. In arbitrary dimensions, the electric and magnetic fields are tensors of different rank. The exception is the four dimensional case where both electric and magnetic fields are rank-1 tensors. In (2+1)D, for instance, the magnetic field is a scalar, while the electric field remains a vector.

To conclude, we point out that to derive the Dirac quantization condition, one usually assumes that the fermion couples minimally with the electromagnetic field, i.e., $p^{\mu}\rightarrow{p^{\mu}-eA^{\mu}}$. However, from the Standard Model Extension perspective, the matter sector can be modified due nonminimal couplings with Lorentz-violating background fields, which may lead to changes in the Dirac condition. We intend to investigate this feature in a future publication \cite{Turcati:2018}.
  
%(Bumblebee electrodynamics)

\acknowledgments
% 
% We would like to thank professors A. Accioly, and J. A. Helayël-Neto for useful comments and discussion. R. T. thanks the Brazilian agency CAPES/PNPD and the Physics Department of the Universidade Federal de Santa Catarina for full support.
We would like to thank professor J. A. Helayël-Neto for useful comments and discussion. R. T. thanks the Physics Department of the Universidade Federal de Santa Catarina for full support. This study was financed in part by the Coordenação de Aperfeiçoamento de Pessoal de Nível Superior - Brasil (CAPES) - Finance Code 001.

\end{document}